\documentclass[aps,prl,reprint,amsmath,amssymb,superscriptaddress,showpacs]{revtex4-1}
\usepackage{hyperref}
\usepackage{amsthm,microtype}

\newcommand{\ket}[1]{\ensuremath{|{#1\rangle}}} 

\newcommand{\braket}[2]{\ensuremath{{\langle #1}|{#2 \rangle}}}
\newcommand{\ketbra}[2]{\ensuremath{|{#1 \rangle}{\langle #2}|}}

\newcommand{\mmt}[1]{#1}
\newcommand{\resp}[1]{\mathcal{#1}}
\newcommand{\D}{\text{d}}
\newtheorem{proposition}{Proposition}
\newtheorem*{definition}{Definition}
\newtheorem*{theorem}{Theorem}
\newtheorem*{lemma}{Lemma}

\begin{document}

\title{Implications of the Pusey--Barrett--Rudolph quantum no-go theorem} 

\author{Maximilian Schlosshauer}

\affiliation{Department of Physics, University of Portland, 5000 North Willamette Boulevard, Portland, Oregon 97203, USA}

\author{Arthur Fine}

\affiliation{Department of Philosophy, University of Washington, Box 353350, Seattle, Washington 98195, USA}


\begin{abstract}
Pusey, Barrett, and Rudolph introduce a new no-go theorem for hidden-variables models of quantum theory. We make precise the class of models targeted and construct equivalent models that evade the theorem. The theorem requires assumptions for models of composite systems, which we examine, determining \emph{compactness} as the weakest assumption needed. On that basis, we demonstrate results of the Bell--Kochen--Specker theorem. Given compactness and the relevant class of models, the theorem can be seen as showing that some measurements on composite systems must have built-in inefficiencies, complicating its testing.
\end{abstract}

\maketitle

Developments in quantum information theory have revived interest in hidden-variables theories \cite{Aaronson:2005:po,*Spekkens:2007:um,*Montina:2011:aa}, most recently focused on a new result, the Pusey--Barrett--Rudolph (PBR) theorem \cite{Pusey:2012:np}. This theorem has emerged as a far-reaching no-go result whose implications are cited as possibly even more dramatic \cite{Leifer:2011:aa} than Bell's theorem \cite{Bell:1964:ep}. One of Einstein's several EPR arguments, for example, can be used to see PBR as offering an alternative route to quantum nonlocality \cite{Harrigan:2010:pl,Pusey:2012:np}. The theorem may also limit promising methods, based on hidden-variables models, for simulating quantum computation classically and efficiently \cite{Pusey:2012:np}. Here we examine the framework of the theorem, including critical assumptions needed to derive it. We characterize the model structures targeted by the theorem and introduce the idea of \emph{built-in inefficiency}, which helps understand the restrictions posed by the theorem and bears on its testability. We identify the weakest assumption on which the theorem rests, and use that to demonstrate the breakdown of additivity associated with the Bell--Kochen--Specker theorem \cite{Bell:1966:ph,Kochen:1967:hu}. PBR understand the significance of their result as undermining the interpretation of the quantum state as ``mere information'' (or ``knowledge'') about the real physical state of a system \cite{Pusey:2012:np}. Here we investigate aspects of the result itself, rather than focusing on interpretive theses proposed on its behalf.

\emph{Hidden variables.---}To provide a context for the PBR theorem, we review the standard approach to modeling a quantum system using hidden variables. For a fixed system, given any state $\ket{\psi}$, a hidden-variables model introduces a probability distribution (density function) $p_\psi$ that is correlated to $\ket{\psi}$ and whose support is a space $\Lambda_\psi$ of ``hidden variables'' $\lambda$, with $\int_{\Lambda_\psi} p_\psi(\lambda) \, \D\lambda = 1$. The $\lambda$'s function to fix outcome probabilities for any measurement of the system. To that end, the model associates with each observable $\mmt{A}$ a \emph{response function} $\resp{A}^\psi(S, \lambda)$ giving the \emph{probability at $\lambda$} that a measurement of $\mmt{A}$, initiated in state $\ket{\psi}$, has an outcome in the set $S$. If $P_\mmt{A}^\psi(S)$ is the Born probability that a measurement of $\mmt{A}$, initiated in state $\ket{\psi}$, yields an outcome in $S$, then we require that 
\begin{equation}\label{eq:matchborn}
P_\mmt{A}^\psi(S) = \int_{\Lambda_\psi} \resp{A}^\psi(S, \lambda) p_\psi(\lambda) \, \D\lambda.
\end{equation}
Thus a successful model retrieves the quantum (Born) probabilities for measurement outcomes of observables as averages over these $\lambda$ probabilities \footnote{One might require more than this, as do the Bell \cite{Bell:1964:ep} and Bell--Kochen--Specker theorems \cite{Bell:1966:ph,Kochen:1967:hu}.}. For simplicity, we choose $\Lambda_\psi$ to have unit measure, $\int_{\Lambda_\psi} \, \D\lambda = 1$. Since the PBR theorem is static, we will not be concerned with the dynamics of state change in hidden-variables models. 
To ensure that every measurement has an outcome, we can require that if $S(\mmt{A})$ is the spectrum of $\mmt{A}$, then
\begin{equation}\label{eq:rfmmt}
\resp{A}^\psi(S(\mmt{A}), \lambda) = 1.
\end{equation}
In general, $\Lambda_\psi$, $p_\psi$, and the response functions can all depend on $\ket{\psi}$. In understanding the implications of the PBR result, it will be important to distinguish \emph{$\psi$-dependent models}, where a response function depends on $\ket{\psi}$, and \emph{$\psi$-independent models}. Some models \emph{require} $\psi$-dependence. For example, if $p_\psi$ is uniform, $p_\psi(\lambda) = 1 \, \forall \lambda \in \Lambda_\psi$, then Eq.~\eqref{eq:matchborn} implies that the model must be $\psi$-dependent for all $\ket{\psi}$. (A model of this type was considered by Bell \cite{Bell:1964:ep}.) 

\emph{PBR theorem.---}Crucially, the PBR theorem, which we now sketch, only concerns \emph{$\psi$-independent} models. For such models, PBR start from a single system and arbitrary, distinct states $\ket{\psi_1}$ and $\ket{\psi_2}$. They suppose that the associated distributions $p_{\psi_1}$ and $p_{\psi_2}$ share at least one hidden variable $\lambda$ in their support (technically, that the overlap of their supports has nonzero measure). They consider a collection of $L$ such systems and assume that each system can be prepared independently in one of the states $\ket{\psi_1}$ and $\ket{\psi_2}$. ($L$ depends on $\braket{\psi_1}{\psi_2}$ in a manner specified by PBR.) This results in $2^L$ possible preparations of the composite system, corresponding to $2^L$ tensor products of the form $\ket{\phi_j} = \ket{\psi_{x_1}} \otimes \ket{\psi_{x_2}} \otimes \cdots \otimes \ket{\psi_{x_L}}$, $x_i \in \{1,2\}$. PBR show how to construct a joint measurement $\mmt{M}$ on the composite system (in general, $\mmt{M}$ is a quantum circuit) such that, under assumptions concerning the composition and independence governing hidden variables $\lambda_\text{c}$ of the \emph{composite}, those $\lambda_\text{c}$ must issue in no result for $\mmt{M}$ \emph{provided its response function is $\psi$-independent}. 

The following lemma is at the heart of the theorem. 

\begin{lemma}
\emph{(PBR)} Consider a discrete observable $\mmt{M}$ with eigenvalues $j=1, 2, \hdots, N$ whose eigenvectors span the space of the system. Suppose there are distinct states $\ket{\phi_j}$ such that $P^{\phi_j}_\mmt{M}\left(\{j\}\right) = 0$. (PBR display cases where $N \ge 4$ and these conditions are satisfied.) Suppose the hidden-variables spaces for these states are not disjoint, so some hidden variable $\lambda_\text{c}$ is contained in every $\Lambda_{\phi_j}$. Then if the model is $\phi_j$-independent so that the response functions satisfy $\resp{M}^{\phi_j}(S,\lambda) \equiv \resp{M}(S,\lambda)$ for all $j$, we get a contradiction. Equation~\eqref{eq:matchborn} implies that
\begin{equation}\label{eq:m}
\resp{M} \left(\{1\},\lambda_\text{c}\right) = \resp{M} \left(\{2\},\lambda_\text{c}\right) = \hdots = \resp{M} \left(\{N\},\lambda_\text{c}\right) = 0.
\end{equation}
So $\resp{M} \left(S(\mmt{M}), \lambda_\text{c}\right) = 0$, contradicting Eq.~\eqref{eq:rfmmt}. 
\end{lemma}

No contradiction arises, however, for  $\phi_j$-dependent models. One can construct such a model for $\mmt{M}$ by using the uniform distribution and letting Eq.~\eqref{eq:matchborn} define the response functions, for all $\lambda$ and $j$, as the Born probabilities themselves: $\resp{M}^{\phi_j}(S,\lambda) = P_\mmt{M}^{\phi_j}(S)$. 

Alternatively, the contradiction is avoidable if we allow for no-shows (measurements with no result) by adding to $S(\mmt{M})$ a conventional null ``value'' $\theta$ to form an \emph{augmented spectrum} $S^+(\mmt{M})$ and require $\resp{M}(S^+(\mmt{M}), \lambda) = 1$ in place of Eq.~\eqref{eq:rfmmt}. The Born probabilities are then recovered conditional on those measurements having an outcome. That is, for any $S \subseteq S(\mmt{M})$, we require
\begin{equation}
P_\mmt{M}^\psi(S) = \frac{\int_{\Lambda_\psi} \resp{M}^\psi(S, \lambda) p_\psi(\lambda) \, \D\lambda}
{\int_{\Lambda_\psi} \resp{M}^\psi\left(S(\mmt{M}), \lambda\right) p_\psi(\lambda) \, \D\lambda},
\end{equation}
rather than Eq.~\eqref{eq:matchborn}. This is the strategy of the ``prism models'' \cite{Pearle:1970:oo,*Fine:1982:hk,*Larsson:1998:ll,*Szabo:2002:lh}, which are local hidden-variables models that accommodate the detection inefficiencies of typical photon experiments testing Bell-like inequalities \cite{Bell:1964:ep}. Then the common $\lambda_\text{c}$ would simply give rise to a built-in inefficiency, equal to $\resp{M} \left( \{\theta\},\lambda_\text{c} \right)$. That is, $\lambda_\text{c}$ predetermines not only outcome probabilities but also whether the system will produce any outcome at all when measured. The term ``built-in'' emphasizes that the no-detection property is intrinsically associated with the system, rather than with ordinary detector errors. Thus the PBR result may be seen as showing how inefficiencies arise as a fundamental property of certain hidden-variables models if the response functions are state independent.

\emph{Mixed versus segregated models.---}Since the PBR theorem is concerned with what happens when hidden variables overlap from one state to another, let us define models as \emph{mixed} if there are distinct $\ket{\psi_1}$ and $\ket{\psi_2}$  whose associated hidden-variables spaces $\Lambda_{\psi_1}$ and $\Lambda_{\psi_2}$ overlap, otherwise call them \emph{segregated} \footnote{We find this terminology less charged than the terms ``$\psi$-epistemic'' and ``$\psi$-ontic'' that PBR adopt from \cite{Harrigan:2010:pl}}. In a segregated model, the connection between hidden variables and states is functional: $\lambda$ corresponds to exactly one pure state $\ket{\psi}$; indeed, each $\lambda \in \Lambda_\psi$ corresponds to that same state and in that sense specifies it. We caution, however, that this correlation should not encourage thinking of the quantum state as a physical property of the system any more than the correspondence between your fingers with the numbers 1 to 10 makes those numbers into physical things. (Correlation is not causation, much less is it physical reduction.) 

A brief sketch of some toy models---mixed and segregated---will prepare us to demonstrate important connections between these two concepts. The states are qubits and the observables are bivalent ($\pm 1$). For simplicity, we consider deterministic models, $\resp{A}^\psi(S,\lambda) \in \{0,1\}$. Then we can write $\resp{A}^\psi(S,\lambda) \equiv \resp{A}^\psi(\lambda)$, which now simply yields the outcome of the $\mmt{A}$ measurement \cite{Fine:1982:hj}. Since our state space is two-dimensional, we ignore complications concerning contextuality arising in dimensions $\ge 3$ \cite{Kochen:1967:hu}.

\emph{Mixed toy model.} No matter what the state, the $\lambda$'s are just numbers in $(0,1]$. For arbitrary $\ket{\psi}$ we represent observable $\mmt{A}$ by a state-dependent response function $\resp{A}^\psi(\lambda)$ with values $\pm 1$, defined as $\resp{A}^\psi(\lambda) = 1$ iff $0 < \lambda \le P^\psi_\mmt{A}\left(\{1\}\right)$. We choose a uniform hidden-variables distribution. Then values of observables are state dependent, but the hidden-variables distribution is not. Trivially, this model returns the Born probabilities, since for every $\ket{\psi}$, $P^\psi_\mmt{A}(1) = \int_0^{P^\psi_\mmt{A}(1)} \D\lambda$. 

\emph{Segregated toy models.} These models mimic the previous one, but here we want the hidden-variables distribution to depend on the state and in such a way, moreover, that no hidden variables overlap from one state to another. Simple modifications readily achieve this. For example, we may assign a different unit interval on the real line to each distinct state and proceed as above for each interval. Or, we may assign each state $\ket{\psi}$ a direction $\hat{\psi}$ on the unit circle and redo the proceeding construction on a unit radius in that direction. Like the original, both models return the quantum statistics; indeed, apart from the particular geometry of segregation, they are structurally identical to the mixed model. We now show that one can always ``unmix'' in the above manner.

\begin{proposition}\label{prop:seg}
Given any mixed model, regardless of any state-dependent response functions, there are structurally equivalent segregated models returning exactly the same statistics.  
\end{proposition}

The trick here is to see that if we make a 1:1 correspondence to some other domain, all probabilistic structure goes through isomorphically when we redefine the distributions and functions using the new names. For example, rename every $\lambda \in \Lambda_\psi$ as a pair $(\lambda, \psi)$ and call the resulting space of pairs $\Lambda_\psi'$. Then whether or not $\Lambda_\psi$ and $\Lambda_\phi$ overlap for $\ket{\psi} \not= \ket{\phi}$, $\Lambda_\psi'$ and $\Lambda_\phi'$ are disjoint. Now redefine the response functions and distributions in terms of the pairs. Let $\resp{A}^\psi(S, \lambda) = {\resp{A}'}^\psi  \left(S, (\lambda,\psi) \right)$ for a state-dependent response function, otherwise set $\resp{A}(S, \lambda) = \resp{A}'\left(S, (\lambda,\psi)\right)$ on each space $\Lambda_\psi'$. If $\Sigma$ is a $p_\psi$-measurable subset of $\Lambda_\psi$, then let $p_\psi\left(\{ \lambda \mid \lambda \in \Sigma \}\right) = p_\psi' \left( \{ (\lambda, \psi) \mid \lambda \in \Sigma \}\right)$ define the distribution $p_\psi'$ and its measurable subsets of $\Lambda_\psi'$. This construction does not change probabilities or expectation values. 

Given our segregated toy models and this simple procedure for segregating any mixed model, it is hard to see physical significance in segregation as such. Perhaps from a computational viewpoint, mixed models may be more efficient in representing information. With that in mind, it is interesting to ask whether a given segregated model can be transformed into an equivalent mixed one. 

\begin{proposition}\label{prop:seg2}
Given any segregated model, there are structurally equivalent mixed models returning exactly the same statistics. 
\end{proposition}

An illustration of this is worked out in \cite{Lewis:2012:az} for a model of Bell's \cite{Bell:1966:ph}. There seems to be a generic way to mix using the following procedure. First tag the segregated elements with new names, in the spirit of the preceding construction. Then map the newly tagged elements from separate spaces onto one. Provided all the spaces have the same cardinality (which one can always take to be the continuum, as in our toy illustrations), this is possible. Finally, use the new names to redefine the distributions and response functions of the mixed structure in the manner of Proposition~\ref{prop:seg}. 

\emph{Understanding PBR: Assumptions and implications.---}The theorem shows the price we may have to pay for a hidden-variables model that is not segregated. We put it this way to make clear that PBR do not show that mixed models are predictively flawed or fail to yield the correct quantum statistics for some observables or states of a given system. Rather, PBR demonstrate a possible difficulty for hidden-variables models in forming composites of identically prepared systems. The difficulty results from $\lambda$'s that breach segregation. In our mixed toy model, this is true for every $\lambda$, likewise for the state-independent model for electron spin constructed in \cite{Kochen:1967:hu}. PBR argue that such shared $\lambda$'s may give rise to hidden variables of the composite that issue in no outcome for certain measurements. Thus the price for using nonsegregated hidden-variables models is that certain measurements on composites \emph{may have} built-in inefficiencies---``may have'' because the demonstrated inefficiency depends critically on two assumptions. 

The first assumption, as we have emphasized, is that the relevant response functions are state independent. This leaves only the distribution of hidden variables to reflect how a measurement ``knows'' the right probabilities from one state to another. While this is an interesting position to investigate, a more common practice in constructing hidden-variables models is to have quantum states provide essential structure that the hidden variables then supplement. This is the path Einstein, whom PBR quote, followed in his one attempt to introduce hidden variables \cite{Belousek:1996:ii,*Holland:2005:po}. Similarly, the de~Broglie--Bohm versions of hidden variables rely on the quantum state, to which they add determinate particle positions or field quantities. 

The second assumption concerns the minefield of how components relate to a composite. PBR assume that hidden variables of components entirely make up hidden variables of the composite, i.e., that given fixed measurement properties, outcome-determining properties of the parts completely make up the corresponding properties of the whole (\cite{Pusey:2012:np}, p.~2). Indeed they assume more. For they conclude that if the probability of overlap for the single systems is $q$, then the probability of overlap for the joint system composed of $L$ such systems is $q^L$. So they take the underlying properties of each subsystem to contribute stochastically independently with respect to the whole. Correlations, however, cannot be ruled out, even if the preparations appear to be independent, because procedures for preparing the individual subsystems may occur together closely in spacetime or share common sources of energy, as well as a common past. On this basis, PBR's independence assumption is questioned by Hall \cite{Hall:2011:xx}, who offers a weaker condition, \emph{compatibility}, as sufficient for a null-measurement result. Compatibility assumes that if $\lambda$ lies in the support of the distributions for each of $\ket{\psi_1}$ and $\ket{\psi_2}$, then it is also in the support of the distributions associated with all product states $\ket{\psi_{x_1}} \otimes \cdots \otimes \ket{\psi_{x_L}}$, $x_i \in \{1,2\}$. Like stochastic independence, this assumes that the very same $\lambda$'s, which govern the response of each component part, also govern the response of the whole. The thought can only be that the parts completely determine the whole. Surely for quantum systems this assumption is very strong, given entanglement and other features of ``quantum wholeness.'' It would be more realistic to allow native variables for composites. One can then weaken the condition necessary for the PBR result to \emph{compactness}, defined as follows.

\begin{definition}
\emph{(Compactness)} If hidden-variables distributions $p_{\psi_1}$ and $p_{\psi_2}$ associated with states $\ket{\psi_1} \not= \ket{\psi_2}$ share at least one $\lambda$ in their support, then there is some $\lambda_\text{c}$ in the support of all the distributions associated with any tensor product of the form $\ket{\psi_{x_1}} \otimes \cdots \otimes \ket{\psi_{x_L}}$, $x_i \in \{1,2\}$.
\end{definition}

We could gloss the principle this way: if the response of every component part is determined by something, then there is something that determines the response of every composite made up of just these parts. This is weaker than compatibility, and does not rely on PBR's stochastic independence. It is nevertheless a strong condition of uniformity (like moving from ``Everyone has a mother'' to ``There is a mother of us all'').  It seems to be the weakest condition that allows a PBR argument to go through. 

With the framework just described in place, we can now formulate the PBR result simply this way. 

\begin{theorem}
\emph{(PBR)} Assuming compactness: if composite hidden-variables models have response functions that are state independent for all tensor products of mixed component states, then some measurements on composites must have built-in inefficiencies.
\end{theorem}

We note that inefficiency based on stochastic independence decreases exponentially; if the compactness condition is used instead, inefficiency for a composite measurement need not reflect any set percentage of mixing among the components at all. We end by displaying the strength of compactness; namely, by showing that, as a simple corollary of the PBR theorem, compactness implies the breakdown of additivity usually associated with the Bell--Kochen--Specker theorem \cite{Bell:1966:ph,Kochen:1967:hu}.

\begin{proposition}\label{prop:additivity}
Assuming compactness and state-independent response functions, even where all observables in a sum commute, the value of the sum may not be equal to the sum of the values of each observable.
\end{proposition}

Consider the measurement $\mmt{M}$ constructed by PBR corresponding to the PBR Lemma above. For $j=1, \hdots, N$ let $\mmt{P}_j = \ketbra{j}{j}$ be the projectors associated with the eigenspaces of $\mmt{M}$. They are pairwise orthogonal, hence commute, and they resolve the identity $I$ since the state space was assumed to be of dimension $N$. Thus
\begin{equation}\label{eq:iden}
\mmt{I} = \mmt{P}_1 + \mmt{P}_2 + \hdots + \mmt{P}_N.
\end{equation}
The hidden-variables model in the PBR Lemma was assumed to be state independent for every $\ket{\phi_j}$. Then Eq.~\eqref{eq:m} implies that a deterministic model would yield $\mmt{P}_j(\lambda_\text{c}) = 0$ for all $j$. But to satisfy requirement Eq.~\eqref{eq:matchborn} on the Born probabilities for the identity, we need $\mmt{I}(\lambda_\text{c}) = 1$. So in Eq.~\eqref{eq:iden} the value of the sum is 1, but the sum of the values is 0. Thus the algebra of commuting observables is not homomorphic to the assignment of values. This is the conclusion of \cite{Bell:1966:ph,Kochen:1967:hu}, often referred to as contextuality. For state-independent models, the assumption of compactness is strong enough to imply it. 

\emph{Conclusions.---}We introduced two important distinctions among hidden-variables models: \emph{mixed} models (overlapping hidden-variables spaces for distinct states) versus \emph{segregated} models, and models with \emph{state-dependent} versus \emph{state-independent} response functions. PBR show that state-independent models of composites formed using systems with mixed models face restrictions. It is vital to see that those restrictions do not imply any difficulty for models of the components themselves. The PBR theorem is not a no-go theorem for the component systems, or for mixed systems generally. 

Moreover, we have shown that the restrictions do not imply that state-independent models for certain systems must generate statistics that violate the Born probabilities. Rather, they imply that some measurements have built-in inefficiencies. This is compatible with obtaining the Born probabilities using the outcomes that are available, which is normal experimental practice in the face of noise and inefficiency. By contrast, Bell's theorem \cite{Bell:1964:ep} demonstrates that certain models must violate the Born statistics if the measurement efficiency exceeds a certain threshold \cite{Cabello:2007:pp}.

In light of Propositions \ref{prop:seg} and \ref{prop:seg2}, segregation and mixing seem to be fungible. Thus if one wants to avoid built-in inefficiency, one can always segregate, without loss, before forming composites. Alternatively, one could allow state dependence.

A remark on testability. Even a large overlap of hidden variables (other than 100\%) may well be buried in the overall inefficiency associated with actual laboratory measurements. Hence, testing the PBR theorem---i.e., testing models for which both state independence and compactness hold---may require looking for experimental signatures other than inefficiencies. One target would be the demonstrated failure of additivity (Proposition \ref{prop:additivity}), which may be amenable to tests of the kind used for the product rule of the Bell--Kochen--Specker theorem \cite{Simon:2000:oa,*Huang:2003:tx,*Kirchmair:2009:aq}.

We thank Matt Leifer and Rob Spekkens for discussions. A.\,F.\ would like to thank John Manchak for introducing him to the PBR theorem.


\begin{thebibliography}{24}%
\makeatletter
\providecommand \@ifxundefined [1]{%
 \@ifx{#1\undefined}
}%
\providecommand \@ifnum [1]{%
 \ifnum #1\expandafter \@firstoftwo
 \else \expandafter \@secondoftwo
 \fi
}%
\providecommand \@ifx [1]{%
 \ifx #1\expandafter \@firstoftwo
 \else \expandafter \@secondoftwo
 \fi
}%
\providecommand \natexlab [1]{#1}%
\providecommand \enquote  [1]{``#1''}%
\providecommand \bibnamefont  [1]{#1}%
\providecommand \bibfnamefont [1]{#1}%
\providecommand \citenamefont [1]{#1}%
\providecommand \href@noop [0]{\@secondoftwo}%
\providecommand \href [0]{\begingroup \@sanitize@url \@href}%
\providecommand \@href[1]{\@@startlink{#1}\@@href}%
\providecommand \@@href[1]{\endgroup#1\@@endlink}%
\providecommand \@sanitize@url [0]{\catcode `\\12\catcode `\$12\catcode
  `\&12\catcode `\#12\catcode `\^12\catcode `\_12\catcode `\%12\relax}%
\providecommand \@@startlink[1]{}%
\providecommand \@@endlink[0]{}%
\providecommand \url  [0]{\begingroup\@sanitize@url \@url }%
\providecommand \@url [1]{\endgroup\@href {#1}{\urlprefix }}%
\providecommand \urlprefix  [0]{URL }%
\providecommand \Eprint [0]{\href }%
\@ifxundefined \urlstyle {%
  \providecommand \doi  [0]{\begingroup \@sanitize@url \@doi}%
  \providecommand \@doi [1]{\endgroup \@@startlink {\doibase
  #1}doi:\discretionary {}{}{}#1\@@endlink }%
}{%
  \providecommand \doi  [0]{doi:\discretionary{}{}{}\begingroup
  \urlstyle{rm}\Url }%
}%
\providecommand \doibase [0]{http://dx.doi.org/}%
\providecommand \Doi [0]{\begingroup \@sanitize@url \@Doi }%
\providecommand \@Doi  [1]{\endgroup\@@startlink{\doibase#1}\@@Doi}%
\providecommand \@@Doi [1]{#1\@@endlink}%
\providecommand \selectlanguage [0]{\@gobble}%
\providecommand \bibinfo  [0]{\@secondoftwo}%
\providecommand \bibfield  [0]{\@secondoftwo}%
\providecommand \translation [1]{[#1]}%
\providecommand \BibitemOpen [0]{}%
\providecommand \bibitemStop [0]{}%
\providecommand \bibitemNoStop [0]{.\EOS\space}%
\providecommand \EOS [0]{\spacefactor3000\relax}%
\providecommand \BibitemShut  [1]{\csname bibitem#1\endcsname}%
\bibitem [{\citenamefont {Aaronson}(2005)}]{Aaronson:2005:po}%
  \BibitemOpen
  \bibfield  {author} {\bibinfo {author} {\bibfnamefont {S.}~\bibnamefont
  {Aaronson}},\ }\href@noop {} {\bibfield  {journal} {\bibinfo  {journal}
  {Phys. Rev. A},\ }\textbf {\bibinfo {volume} {71}},\ \bibinfo {pages}
  {032325} (\bibinfo {year} {2005})}\BibitemShut {NoStop}%
\bibitem [{\citenamefont {Spekkens}(2007)}]{Spekkens:2007:um}%
  \BibitemOpen
  \bibfield  {author} {\bibinfo {author} {\bibfnamefont {R.~W.}\ \bibnamefont
  {Spekkens}},\ }\href@noop {} {\bibfield  {journal} {\bibinfo  {journal}
  {Phys. Rev. A},\ }\textbf {\bibinfo {volume} {75}},\ \bibinfo {pages}
  {032110} (\bibinfo {year} {2007})}\BibitemShut {NoStop}%
\bibitem [{\citenamefont {Montina}(2011)}]{Montina:2011:aa}%
  \BibitemOpen
  \bibfield  {author} {\bibinfo {author} {\bibfnamefont {A.}~\bibnamefont
  {Montina}},\ }\href@noop {} {\bibfield  {journal} {\bibinfo  {journal} {Phys.
  Rev. A},\ }\textbf {\bibinfo {volume} {83}},\ \bibinfo {pages} {032107}
  (\bibinfo {year} {2011})}\BibitemShut {NoStop}%
\bibitem [{\citenamefont {Pusey}\ \emph {et~al.}(2012)\citenamefont {Pusey},
  \citenamefont {Barrett},\ and\ \citenamefont {Rudolph}}]{Pusey:2012:np}%
  \BibitemOpen
  \bibfield  {author} {\bibinfo {author} {\bibfnamefont {M.~F.}\ \bibnamefont
  {Pusey}}, \bibinfo {author} {\bibfnamefont {J.}~\bibnamefont {Barrett}}, \
  and\ \bibinfo {author} {\bibfnamefont {T.}~\bibnamefont {Rudolph}},\
  }\href@noop {} {\bibfield  {journal} {\bibinfo  {journal} {Nature Physics},\
  }\textbf {\bibinfo {volume} {8}},\ \bibinfo {pages} {476} (\bibinfo {year}
  {2012})}\BibitemShut {NoStop}%
\bibitem [{\citenamefont {Leifer}(2011)}]{Leifer:2011:aa}%
  \BibitemOpen
  \bibfield  {author} {\bibinfo {author} {\bibfnamefont {M.}~\bibnamefont
  {Leifer}},\ }\href@noop {} {\bibfield  {journal} {\bibinfo  {journal} {The
  Quantum Times},\ }\textbf {\bibinfo {volume} {6}},\ \bibinfo {pages} {1}
  (\bibinfo {year} {2011})}\BibitemShut {NoStop}%
\bibitem [{\citenamefont {Bell}(1964)}]{Bell:1964:ep}%
  \BibitemOpen
  \bibfield  {author} {\bibinfo {author} {\bibfnamefont {J.~S.}\ \bibnamefont
  {Bell}},\ }\href@noop {} {\bibfield  {journal} {\bibinfo  {journal}
  {Physics},\ }\textbf {\bibinfo {volume} {1}},\ \bibinfo {pages} {195}
  (\bibinfo {year} {1964})}\BibitemShut {NoStop}%
\bibitem [{\citenamefont {Harrigan}\ and\ \citenamefont
  {Spekkens}()}]{Harrigan:2010:pl}%
  \BibitemOpen
  \bibfield  {author} {\bibinfo {author} {\bibfnamefont {N.}~\bibnamefont
  {Harrigan}}\ and\ \bibinfo {author} {\bibfnamefont {R.~W.}\ \bibnamefont
  {Spekkens}},\ }\href@noop {} {\bibfield  {journal} {\bibinfo  {journal}
  {Found. Phys.},\ }\textbf {\bibinfo {volume} {40}},\ \bibinfo {pages}
  {125}}\BibitemShut {NoStop}%
\bibitem [{\citenamefont {Bell}(1966)}]{Bell:1966:ph}%
  \BibitemOpen
  \bibfield  {author} {\bibinfo {author} {\bibfnamefont {J.~S.}\ \bibnamefont
  {Bell}},\ }\href@noop {} {\bibfield  {journal} {\bibinfo  {journal} {Rev.
  Mod. Phys.},\ }\textbf {\bibinfo {volume} {38}},\ \bibinfo {pages} {447}
  (\bibinfo {year} {1966})}\BibitemShut {NoStop}%
\bibitem [{\citenamefont {Kochen}\ and\ \citenamefont
  {Specker}(1967)}]{Kochen:1967:hu}%
  \BibitemOpen
  \bibfield  {author} {\bibinfo {author} {\bibfnamefont {S.}~\bibnamefont
  {Kochen}}\ and\ \bibinfo {author} {\bibfnamefont {E.}~\bibnamefont
  {Specker}},\ }\href@noop {} {\bibfield  {journal} {\bibinfo  {journal} {J.
  Math. Mech.},\ }\textbf {\bibinfo {volume} {17}},\ \bibinfo {pages} {59}
  (\bibinfo {year} {1967})}\BibitemShut {NoStop}%
\bibitem [{Note1()}]{Note1}%
  \BibitemOpen
  \bibinfo {note} {One might require more than this, as do the Bell \cite
  {Bell:1964:ep} and Bell--Kochen--Specker theorems \cite
  {Bell:1966:ph,Kochen:1967:hu}.}\BibitemShut {Stop}%
\bibitem [{\citenamefont {Pearle}(1970)}]{Pearle:1970:oo}%
  \BibitemOpen
  \bibfield  {author} {\bibinfo {author} {\bibfnamefont {P.~M.}\ \bibnamefont
  {Pearle}},\ }\href@noop {} {\bibfield  {journal} {\bibinfo  {journal} {Phys.
  Rev. D},\ }\textbf {\bibinfo {volume} {2}},\ \bibinfo {pages} {1418}
  (\bibinfo {year} {1970})}\BibitemShut {NoStop}%
\bibitem [{\citenamefont {Fine}(1982){\natexlab{a}}}]{Fine:1982:hk}%
  \BibitemOpen
  \bibfield  {author} {\bibinfo {author} {\bibfnamefont {A.}~\bibnamefont
  {Fine}},\ }\href@noop {} {\bibfield  {journal} {\bibinfo  {journal}
  {Synthese},\ }\textbf {\bibinfo {volume} {50}},\ \bibinfo {pages} {279}
  (\bibinfo {year} {1982}{\natexlab{a}})}\BibitemShut {NoStop}%
\bibitem [{\citenamefont {Larsson}(1998)}]{Larsson:1998:ll}%
  \BibitemOpen
  \bibfield  {author} {\bibinfo {author} {\bibfnamefont {J.-{\AA}.}\
  \bibnamefont {Larsson}},\ }\href@noop {} {\bibfield  {journal} {\bibinfo
  {journal} {Phys. Rev. A},\ }\textbf {\bibinfo {volume} {57}},\ \bibinfo
  {pages} {R3145} (\bibinfo {year} {1998})}\BibitemShut {NoStop}%
\bibitem [{\citenamefont {Szab{\'o}}\ and\ \citenamefont
  {Fine}(2002)}]{Szabo:2002:lh}%
  \BibitemOpen
  \bibfield  {author} {\bibinfo {author} {\bibfnamefont {L.~E.}\ \bibnamefont
  {Szab{\'o}}}\ and\ \bibinfo {author} {\bibfnamefont {A.}~\bibnamefont
  {Fine}},\ }\href@noop {} {\bibfield  {journal} {\bibinfo  {journal} {Phys.
  Lett. A},\ }\textbf {\bibinfo {volume} {295}},\ \bibinfo {pages} {229}
  (\bibinfo {year} {2002})}\BibitemShut {NoStop}%
\bibitem [{Note2()}]{Note2}%
  \BibitemOpen
  \bibinfo {note} {We find this terminology less charged than the terms ``$\psi
  $-epistemic'' and ``$\psi $-ontic'' that PBR adopt from \cite
  {Harrigan:2010:pl}}\BibitemShut {NoStop}%
\bibitem [{\citenamefont {Fine}(1982){\natexlab{b}}}]{Fine:1982:hj}%
  \BibitemOpen
  \bibfield  {author} {\bibinfo {author} {\bibfnamefont {A.}~\bibnamefont
  {Fine}},\ }\href@noop {} {\bibfield  {journal} {\bibinfo  {journal} {Phys.
  Rev. Lett.},\ }\textbf {\bibinfo {volume} {48}},\ \bibinfo {pages} {291}
  (\bibinfo {year} {1982}{\natexlab{b}})}\BibitemShut {NoStop}%
\bibitem [{\citenamefont {Lewis}\ \emph {et~al.}()\citenamefont {Lewis},
  \citenamefont {Jennings}, \citenamefont {Barrett},\ and\ \citenamefont
  {Rudolph}}]{Lewis:2012:az}%
  \BibitemOpen
  \bibfield  {author} {\bibinfo {author} {\bibfnamefont {P.~G.}\ \bibnamefont
  {Lewis}}, \bibinfo {author} {\bibfnamefont {D.}~\bibnamefont {Jennings}},
  \bibinfo {author} {\bibfnamefont {J.}~\bibnamefont {Barrett}}, \ and\
  \bibinfo {author} {\bibfnamefont {T.}~\bibnamefont {Rudolph}},\ }\href@noop
  {} {}\Eprint {http://arxiv.org/abs/arXiv:1201.6554v1 [quant-ph]}
  {arXiv:1201.6554v1 [quant-ph]} \BibitemShut {NoStop}%
\bibitem [{\citenamefont {Belousek}(1996)}]{Belousek:1996:ii}%
  \BibitemOpen
  \bibfield  {author} {\bibinfo {author} {\bibfnamefont {D.~W.}\ \bibnamefont
  {Belousek}},\ }\href@noop {} {\bibfield  {journal} {\bibinfo  {journal}
  {Stud. Hist. Phil. Mod. Phys.},\ }\textbf {\bibinfo {volume} {27}},\ \bibinfo
  {pages} {437} (\bibinfo {year} {1996})}\BibitemShut {NoStop}%
\bibitem [{\citenamefont {Holland}(2005)}]{Holland:2005:po}%
  \BibitemOpen
  \bibfield  {author} {\bibinfo {author} {\bibfnamefont {P.}~\bibnamefont
  {Holland}},\ }\href@noop {} {\bibfield  {journal} {\bibinfo  {journal}
  {Found. Phys.},\ }\textbf {\bibinfo {volume} {35}},\ \bibinfo {pages} {177}
  (\bibinfo {year} {2005})}\BibitemShut {NoStop}%
\bibitem [{\citenamefont {Hall}()}]{Hall:2011:xx}%
  \BibitemOpen
  \bibfield  {author} {\bibinfo {author} {\bibfnamefont {M.~J.~W.}\
  \bibnamefont {Hall}},\ }\href@noop {} {}\Eprint
  {http://arxiv.org/abs/arXiv:1111.6304v1 [quant-ph]} {arXiv:1111.6304v1
  [quant-ph]} \BibitemShut {NoStop}%
\bibitem [{\citenamefont {Cabello}\ and\ \citenamefont
  {Larsson}(2007)}]{Cabello:2007:pp}%
  \BibitemOpen
  \bibfield  {author} {\bibinfo {author} {\bibfnamefont {A.}~\bibnamefont
  {Cabello}}\ and\ \bibinfo {author} {\bibfnamefont {J.-{\AA}.}\ \bibnamefont
  {Larsson}},\ }\href@noop {} {\bibfield  {journal} {\bibinfo  {journal} {Phys.
  Rev. Lett.},\ }\textbf {\bibinfo {volume} {98}},\ \bibinfo {pages} {220402}
  (\bibinfo {year} {2007})}\BibitemShut {NoStop}%
\bibitem [{\citenamefont {Simon}\ \emph {et~al.}(2000)\citenamefont {Simon},
  \citenamefont {{\.{Z}}ukowski}, \citenamefont {Weinfurter},\ and\
  \citenamefont {Zeilinger}}]{Simon:2000:oa}%
  \BibitemOpen
  \bibfield  {author} {\bibinfo {author} {\bibfnamefont {C.}~\bibnamefont
  {Simon}}, \bibinfo {author} {\bibfnamefont {M.}~\bibnamefont
  {{\.{Z}}ukowski}}, \bibinfo {author} {\bibfnamefont {H.}~\bibnamefont
  {Weinfurter}}, \ and\ \bibinfo {author} {\bibfnamefont {A.}~\bibnamefont
  {Zeilinger}},\ }\href@noop {} {\bibfield  {journal} {\bibinfo  {journal}
  {Phys. Rev. Lett.},\ }\textbf {\bibinfo {volume} {85}},\ \bibinfo {pages}
  {1783} (\bibinfo {year} {2000})}\BibitemShut {NoStop}%
\bibitem [{\citenamefont {Huang}\ \emph {et~al.}(2003)\citenamefont {Huang},
  \citenamefont {Li}, \citenamefont {Zhang}, \citenamefont {Pan},\ and\
  \citenamefont {Guo}}]{Huang:2003:tx}%
  \BibitemOpen
  \bibfield  {author} {\bibinfo {author} {\bibfnamefont {Y.-F.}\ \bibnamefont
  {Huang}}, \bibinfo {author} {\bibfnamefont {C.-F.}\ \bibnamefont {Li}},
  \bibinfo {author} {\bibfnamefont {Y.-S.}\ \bibnamefont {Zhang}}, \bibinfo
  {author} {\bibfnamefont {J.-W.}\ \bibnamefont {Pan}}, \ and\ \bibinfo
  {author} {\bibfnamefont {G.-C.}\ \bibnamefont {Guo}},\ }\href@noop {}
  {\bibfield  {journal} {\bibinfo  {journal} {Phys. Rev. Lett.},\ }\textbf
  {\bibinfo {volume} {90}},\ \bibinfo {pages} {250401} (\bibinfo {year}
  {2003})}\BibitemShut {NoStop}%
\bibitem [{\citenamefont {Kirchmair}\ \emph {et~al.}(2009)\citenamefont
  {Kirchmair}, \citenamefont {Z{\"a}hringer}, \citenamefont {Gerritsma},
  \citenamefont {Kleinmann}, \citenamefont {G{\"u}hne}, \citenamefont
  {Cabello}, \citenamefont {Blatt},\ and\ \citenamefont
  {Roos}}]{Kirchmair:2009:aq}%
  \BibitemOpen
  \bibfield  {author} {\bibinfo {author} {\bibfnamefont {G.}~\bibnamefont
  {Kirchmair}}, \bibinfo {author} {\bibfnamefont {F.}~\bibnamefont
  {Z{\"a}hringer}}, \bibinfo {author} {\bibfnamefont {R.}~\bibnamefont
  {Gerritsma}}, \bibinfo {author} {\bibfnamefont {M.}~\bibnamefont
  {Kleinmann}}, \bibinfo {author} {\bibfnamefont {O.}~\bibnamefont
  {G{\"u}hne}}, \bibinfo {author} {\bibfnamefont {A.}~\bibnamefont {Cabello}},
  \bibinfo {author} {\bibfnamefont {R.}~\bibnamefont {Blatt}}, \ and\ \bibinfo
  {author} {\bibfnamefont {C.~F.}\ \bibnamefont {Roos}},\ }\href@noop {}
  {\bibfield  {journal} {\bibinfo  {journal} {Nature},\ }\textbf {\bibinfo
  {volume} {460}},\ \bibinfo {pages} {494} (\bibinfo {year}
  {2009})}\BibitemShut {NoStop}%
\end{thebibliography}

%

\end{document}